\documentclass{article}
\usepackage{spconf,amsmath,graphicx,hyperref}
\usepackage{algorithm} \usepackage{algpseudocode}
\usepackage{hyperref}
\usepackage{spconf}
\usepackage[skip=2pt]{caption}
\usepackage[utf8]{inputenc}
\usepackage[T1]{fontenc}
\usepackage{textcomp}       
\usepackage{algorithmicx, algpseudocode}
\usepackage{enumitem}
\usepackage{balance}

\title{Time-Frequency Analysis of Non-Uniformly Sampled Signals via \\ sample density adaptation}
\name{Ashwini Kulkarni and Santosh Nannuru}
\address{Signal Processing and Communication Research Center, IIIT Hyderabad, India}
\begin{document}
%
\maketitle
\ninept 

\begin{abstract}

The analysis of non-stationary signals in non-uniformly sampled data is a challenging task. Time-integrated methods, such as the generalised Lomb-Scargle (GLS) periodogram, provide a robust statistical assessment of persistent periodicities but are insensitive to transient events. Conversely, existing time-frequency methods often rely on fixed-duration windows or interpolation, which can be suboptimal for non-uniform data. 
We introduce the non-uniform Stockwell-transform (NUST), a time-frequency framework that applies a localized density adaptive spectral analysis directly to non-uniformly sampled data. 
NUST employs a doubly adaptive window that adjusts its width based on both frequency and local data density, providing detailed time-frequency information for both transient and persistent signals.
We validate the NUST on numerous non-uniformly sampled synthetic signals, demonstrating its superior time-localization performance compared to GLS.
Furthermore, we apply NUST to HARPS radial velocity data of the multi-planetary system HD 10180, successfully distinguishing coherent planetary signals from stellar activity. 
\end{abstract}
\section{Introduction}
\label{sec:intro}

The analysis of time-series data underpins discovery in many scientific and engineering fields. 
While many signal processing techniques are built for uniformly sampled data, real-world applications often produce a different reality.
For example, data from astronomical surveys \cite{vanderplas2018understanding}, geophysical monitoring \cite{ghil2002advanced}, and biomedical sensors \cite{laguna1998power} are non-uniformly sampled, sparse, or riddled with gaps (i.e., $t_{i+1} - t_i \neq \text{constant}$).
This non-uniformity makes time-frequency analysis challenging, specifically for non-stationary signals.

A straightforward approach is to first interpolate the data onto a uniform grid and then apply standard techniques like the short-time Fourier transform (STFT) \cite{cohen1995time}, Stockwell transform \cite{stockwell2002localization} or wavelet transform \cite{mallat1999wavelet}. However, interpolation can introduce significant artifacts \cite{eyer1999variable}. Although variants of STFT that use the non-uniform fast Fourier transform \cite{fessler2003nonuniform} can operate on non-uniform samples, they are limited by a fixed size analysis window.

A variety of methods have been developed to operate directly on non-uniform data. These are typically time-integrated, producing a single global spectrum, like the generalised Lomb-Scargle periodogram \cite{lomb1976least, zechmeister2009generalised}, which uses a least-squares fit of sinusoids to the data. 
While effective, the time-integrated nature of these methods makes them blind to transient or non-stationary phenomena, such as a signal appearing for a limited duration or drifting in frequency.


To provide temporal localisation, methods that employ a windowed or time-frequency approach have been developed. The Lomb-Welch periodogram \cite{thong2004lomb} applies the Lomb-Scargle analysis to short, overlapping segments of equal time duration and averages the resulting spectra to produce a variance-reduced spectrogram.
Other techniques like the weighted wavelet Z-transform (WWZ) \cite{foster1996wavelets} use wavelets to achieve time-frequency localization. Least-squares wavelet analysis (LSWA) \cite{Ghaderpour2017LeastSquaresWA} improves on this by using a frequency-scaled window to overcome the limitations of a fixed resolution.
Although these methods provide crucial time localization, they overlook the local sample density, often yielding unreliable estimates, especially in sparsely sampled regions.


In this paper, we introduce the non-uniform S-transform (NUST), a time-frequency framework designed to address the need for sample density adaptation.
It synthesizes the statistical robustness of the GLS periodogram \cite{zechmeister2009generalised} with the adaptive multi-resolution principles of the S-transform \cite{stockwell2002localization}.
The S-transform's Fourier kernel is replaced with a localised GLS analysis. In this scheme, the Gaussian window of the S-transform is doubly adaptive, with its width governed jointly by the signal frequency and the local sample density.
We present the mathematical formulation of NUST, including its sample density-adaptive components, and validate its performance against traditional methods on challenging synthetic and real-world signals.

\section{Background}
\label{sec:format}

\subsection{Time-frequency analysis for uniformly sampled data}
Analyzing signals whose frequency content is not constant, such as weak, transient, or evolving signals requires methods that provide temporal localization.
The foundational approach is the short-time Fourier transform (STFT), which applies the standard Fourier transform to windowed segments of a signal.
While providing time-frequency information, the fixed-width window imposes a rigid resolution trade-off, limiting its ability to simultaneously resolve short, high-frequency bursts and long-duration, low-frequency oscillations.
The Stockwell transform (S-transform) provides an elegant solution to this limitation. It employs a Gaussian window whose width is dynamically scaled to be inversely proportional to the frequency, offering a powerful multi-resolution analysis. The continuous S-transform of a signal $x(t)$ is defined as :
\begin{equation}
    S(\tau, f) = \int_{-\infty}^{\infty} x(t) \frac{|f|}{\sqrt{2\pi}} e^{-\frac{(t-\tau)^2 f^2}{2}} e^{-i2\pi f t} dt
    \label{eq:s_transform}
\end{equation}
The standard deviation of the window, $\sigma$, is the key to its adaptivity, \(
    \sigma = \frac{1}{|f|}
    \label{eq:s_transform_sigma}\).
However, its reliance on the Fourier kernel means it is fundamentally defined only for uniformly sampled data, creating a significant barrier for many practical applications.

\subsection{Spectral analysis of non-uniformly sampled data: The GLS periodogram}
A statistical tool for detecting periodic sinusoidal signals in non-uniformly sampled data is the generalized Lomb-Scargle (GLS) periodogram. The pseudocode for GLS is outlined in Algorithm \ref{alg:gls_formal}. 

For each trial frequency $f$, GLS performs a least-squares fit of the sinusoidal model \(y(t) = a\cos(2\pi ft) + b\sin(2\pi ft) + c\) to the data.
The method seeks to minimize the weighted sum of squared residuals ($\chi^2$), where weights $w_i$ are typically the inverse measurement variance ($1/\sigma_i^2$). The power of the GLS periodogram, $p(f)$, is a normalized measure of the goodness-of-fit, defined as the fractional reduction in $\chi^2$ achieved by the sinusoidal model compared to a constant-offset model ($\chi^2_0$),
\begin{equation}
    p(f) = \frac{\chi_0^2 - \chi^2_{\text{fit}}(f)}{\chi_0^2} \,.
    \label{eq:gls_power}
\end{equation}
However, since GLS is a time-integrated method, it cannot capture any temporal variations in a signal, such as transient behaviour or frequency drift, which is its key limitation in the analysis of non-stationary signals. 

\begin{algorithm}[H] 
\caption{Generalized Lomb-Scargle (GLS) Periodogram}
\label{alg:gls_formal}
\begin{algorithmic}[1]
    \Require Time series data $\{(t_i, y_i, \sigma_i)\}_{i=1}^N$, frequencies $\{f_k\}$
    \Procedure{GLS}{$\{t_i\}, \{y_i\}, \{\sigma_i\}, \{f_k\}$}
        \State Initialize weights $w_i \gets 1/\sigma_i^2$, mean $\bar{y} = \tfrac{\sum w_i y_i}{\sum w_i}$, and $\chi_0^2 \gets \sum w_i (y_i - \bar{y})^2$
        \ForAll{$f \in \{f_k\}$}
            \State \textbf{Least-squares fit of the model:}
            \State \(y(t) = a\cos(2\pi f t) + b\sin(2\pi f t) + c\)
            \State Find coefficients $a, b, c$ that minimize
            \State \(\chi^2(f) = \sum_{i=1}^N w_i 
            \big[y_i - y(t_i)\big]^2\)
            \State Let minimized value be $\chi_{\text{fit}}^2(f)$
            \State Compute normalized power: $p(f) \gets \tfrac{\chi_0^2 - \chi_{\text{fit}}^2(f)}{\chi_0^2}$
        \EndFor
    \EndProcedure
\end{algorithmic}
\end{algorithm}

\section{Non-uniform Stockwell transform}
\label{sec:nust}
We introduce the non-uniform S-transform (NUST), a novel time-frequency framework for analyzing non-uniformly sampled data. It uniquely combines S-Transform and GLS periodogram to overcome their respective limitations. 
At its core, the method applies the GLS periodogram, not to the entire dataset, but to data points contained within a sliding temporal window. Thus, NUST retains the statistical robustness of the GLS while introducing the critical element of time localization.

The primary innovation of the NUST lies in its doubly adaptive windowing scheme. The window width is modulated by two key factors, frequency and data density. Following the principle of the S-Transform, the window narrows for high frequencies to achieve high temporal resolution and widens for low frequencies to improve frequency resolution. The window width is also scaled according to the local density of the data. It expands in sparsely sampled regions to gather sufficient information for a stable GLS estimate and contracts in densely sampled regions to provide finer temporal detail. The implementation of this framework rests on two key components: data density estimation and a localized GLS core.

\subsection{Data density estimation and sample-adaptive windowing}
The standard S-Transform's window width, $\sigma \propto 1/|f|$, is optimal only under the implicit assumption of uniform data density. 
This assumption becomes problematic for non-uniform data. During a high-frequency search, the analysis window can become very narrow, and if this occurs in a sparsely sampled region, the GLS fit will not be robust because it is performed on too few data points. 
The analysis window should dynamically adjust its temporal support, narrowing where data is dense to capture fine temporal details, and widening where data is sparse to gather sufficient information for a stable frequency estimate.

To quantify the local data density, we employ kernel density estimation (KDE), a non-parametric method for estimating a density function, $\rho(t)$, from the discrete set of time samples $\{t_i\}$. The density estimate at time $t$ is given by,
\begin{equation}
    \rho(t) = \frac{1}{Nh} \sum_{i=1}^{N} K\left(\frac{t - t_i}{h}\right)
    \label{eq:kde_density}
\end{equation}
where $N$ is the number of samples, $K(\cdot)$ is a kernel function, and $h$ is the bandwidth. 
The bandwidth hyperparameter $h$ controls the shape of the density estimate. A small value of $h$ results in a bumpy density curve. In contrast, a large value of $h$ creates a smoother density estimate. While various kernels are available \cite{wkeglarczyk2018kernel}, the Gaussian kernel is often preferred and is used in our study as well.


With this continuous density estimate, we define the standard deviation of the NUST's Gaussian analysis window to be a function of both the analysis time $\tau$ and the frequency $f$,
\begin{equation}
    \sigma_{\text{adaptive}}(\tau, f) = \frac{\alpha}{|f| \cdot [{\rho}(\tau)]^{\gamma}} \,.
    \label{eq:doubly_adaptive_window}
\end{equation}
Here, $\alpha$ is a scaling hyperparameter that sets the baseline window width. The parameter $\gamma$ is a sensitivity hyperparameter that controls how strongly the window width reacts to changes in local data density. 



\begin{figure}[t] 
    \centering
    \includegraphics[width=\columnwidth]{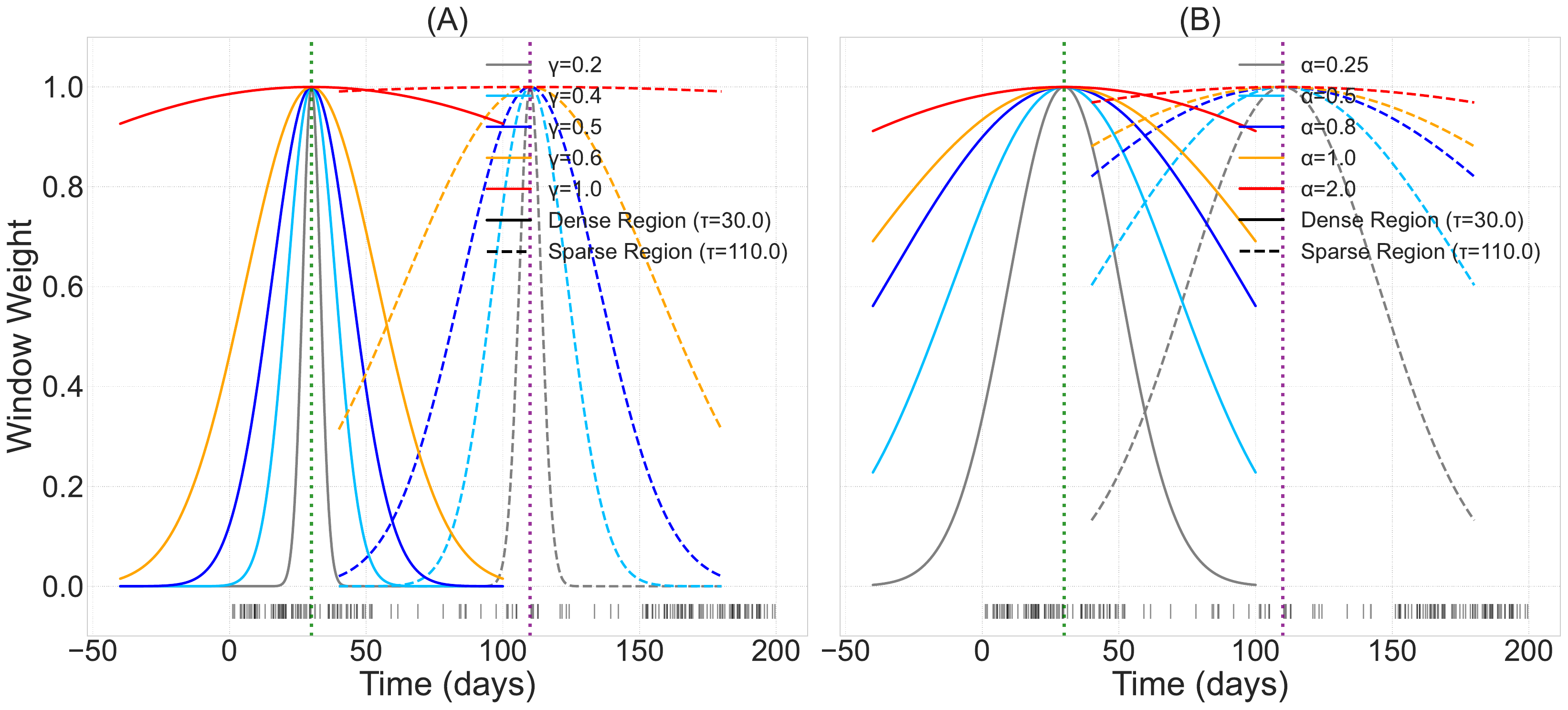}
    \caption{Adaptive window weights at two time instances, $\tau=30$ (solid) at dense and $\tau=110$ (dash) sparse regions, marked by dotted vertical lines. Panel (A) uses $h = 20$, $\alpha = 0.18$ with varying $\gamma$, while Panel (B) uses $h = 20$, $\gamma = 0.5$ with varying $\alpha$.}
    \vspace{-6pt}
    \label{fig:hypermecha}
\end{figure}

\subsection{The Localised GLS Core and Final NUST Power}

The second step in the NUST framework is the computation of spectral power at various points in the time-frequency plane using a localized GLS analysis. 
The doubly adaptive Gaussian window for the time-frequency point $(\tau, f)$ is given by,
\begin{equation}
    g_{\text{adaptive}}(t; \tau, f) = \exp\left(-\frac{(t - \tau)^2}{2\sigma_{\text{adaptive}}(\tau, f)^2}\right) \,.
    \label{eq:gadaptive}
\end{equation}
Some examples of Gaussian windows designed using \eqref{eq:doubly_adaptive_window} and \eqref{eq:gadaptive} are shown in \autoref{fig:hypermecha}.
This adaptive window is then used to compute a total, combined weight, $W_i$, for each input data point $(t_i, y_i)$. This weight is the product of the intrinsic measurement weight $w_{\text{meas},i} = 1/\sigma_i^2$, where $\sigma_i$ is the measurement uncertainty, 
and the adaptive Gaussian window $g_{\text{adaptive}}$,
\begin{equation}
    W_i(\tau, f) = w_{\text{meas},i} \cdot g_{\text{adaptive}}(t_i; \tau, f) \,.
    \label{eq:combined_weight}
\end{equation}
This combined weight ensures that the subsequent fit is simultaneously sensitive to measurement uncertainty, the local sampling structure, and the specific time-frequency coordinate being analyzed.
Using these new weights $\{W_i(\tau,f)\}$, a least-squares fit is performed for a sinusoidal model of the form $y(t) = a\cos(2\pi f t) + b\sin(2\pi f t) + c$,
as outlined in Algorithm \ref{alg:gls_formal}. The output of the NUST at the point $(\tau,f)$, denoted $S_{\text{NUST}}(\tau,f)$, is the statistical power, $p(\tau,f)$,
\begin{equation}
    S_{\text{NUST}}(\tau,f) \equiv p(\tau,f) = \frac{\chi^2_0 - \chi^2_{\text{fit}}(f)}{\chi^2_0} \,.
    \label{eq:nust_power}
\end{equation}
Computationally, it is calculated using the explicit GLS formula, adapted with our localized weights $W_i$.

\section{Experimental Validation}
\label{sec:typestyle}
\subsection{Experimental setup}
A suite of four synthetic signals are designed to test NUST's ability to characterize signals that are transient, localized, or vary quickly with frequency. The four signals (\autoref{fig:timeseries}) are described below.
\begin{itemize}[left=0pt]
    \setlength\itemsep{0em}   
    \setlength\parskip{0em}
    \setlength\itemindent{0pt}
    \setlength\labelsep{0.5em}
    \item \textbf{Signal 1 (Multi-transient):} A signal consisting of three time-localized sinusoidal components: a $0.07\,\text{cycles/day}$ sinusoid of amplitude $0.9$ active between time $20$--$70\text{days}$, a $0.25\,\text{c/d}$ sinusoid of amplitude $1.0$ active between time $100$--$140\text{d}$, and a $0.15\,\text{c/d}$ sinusoid of amplitude $0.8$ active between time $180$--$210\text{d}$.
    \item \textbf{Signal 2 (Transient chirp):} A chirp signal active between $50$--$150\text{d}$ with constant amplitude $1.0$ and frequency increasing linearly from $0.1\,\text{c/d}$ to $0.2\,\text{c/d}$.
    \item \textbf{Signal 3 (Transient burst):} A localized burst signal consisting of a $0.4\,\text{c/d}$ sinusoid of amplitude $1.0$, active during $95$--$105\text{d}$.
    \item \textbf{Signal 4 (Central gap signal):} A sinusoid of frequency $0.08\,\text{c/d}$ and amplitude 1.0 is present for the entire duration except for a gap between $80$--$160\text{d}$, where the signal is absent.
\end{itemize} 
The non-uniform time samples were drawn from a uniform distribution spanning the signal duration. For Signal $3$ fewer points were drawn during the burst phase to deliberately create regions of sparse observational coverage.

\begin{figure}[h] 
    \centering
    \includegraphics[width=\columnwidth]{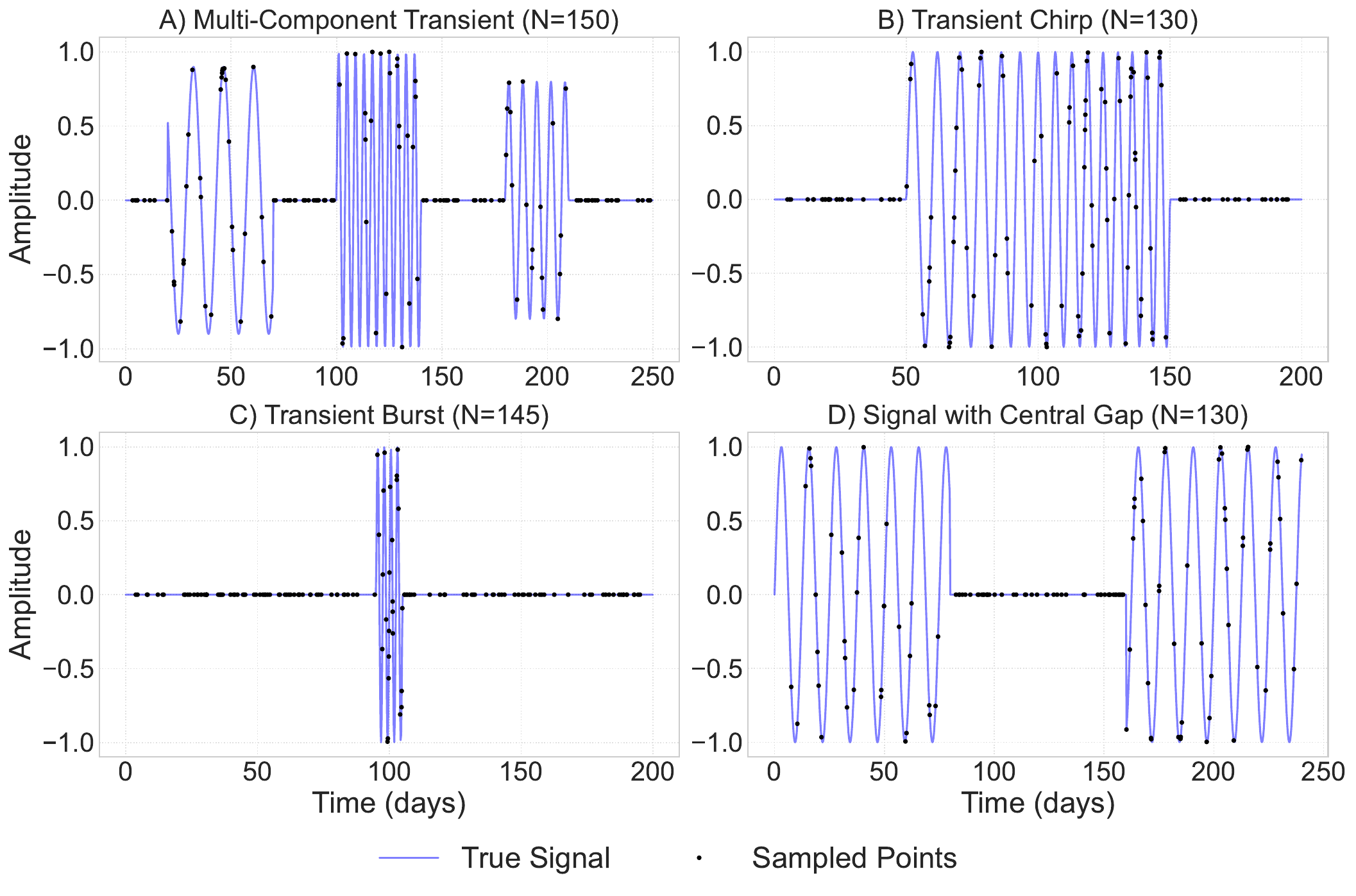}
    \caption{Illustration of the four test signals used in our study. The continuous signal is in blue, and the non-uniformly sampled points are marked in black. (A) Multi-transient signal, (B) Transient chirp, (C) Transient burst, and (D) Central gap signal.
    }
    \vspace{-6pt}
    \label{fig:timeseries}
\end{figure}
\begin{figure}[h]
    \centering
    \includegraphics[width=\columnwidth]{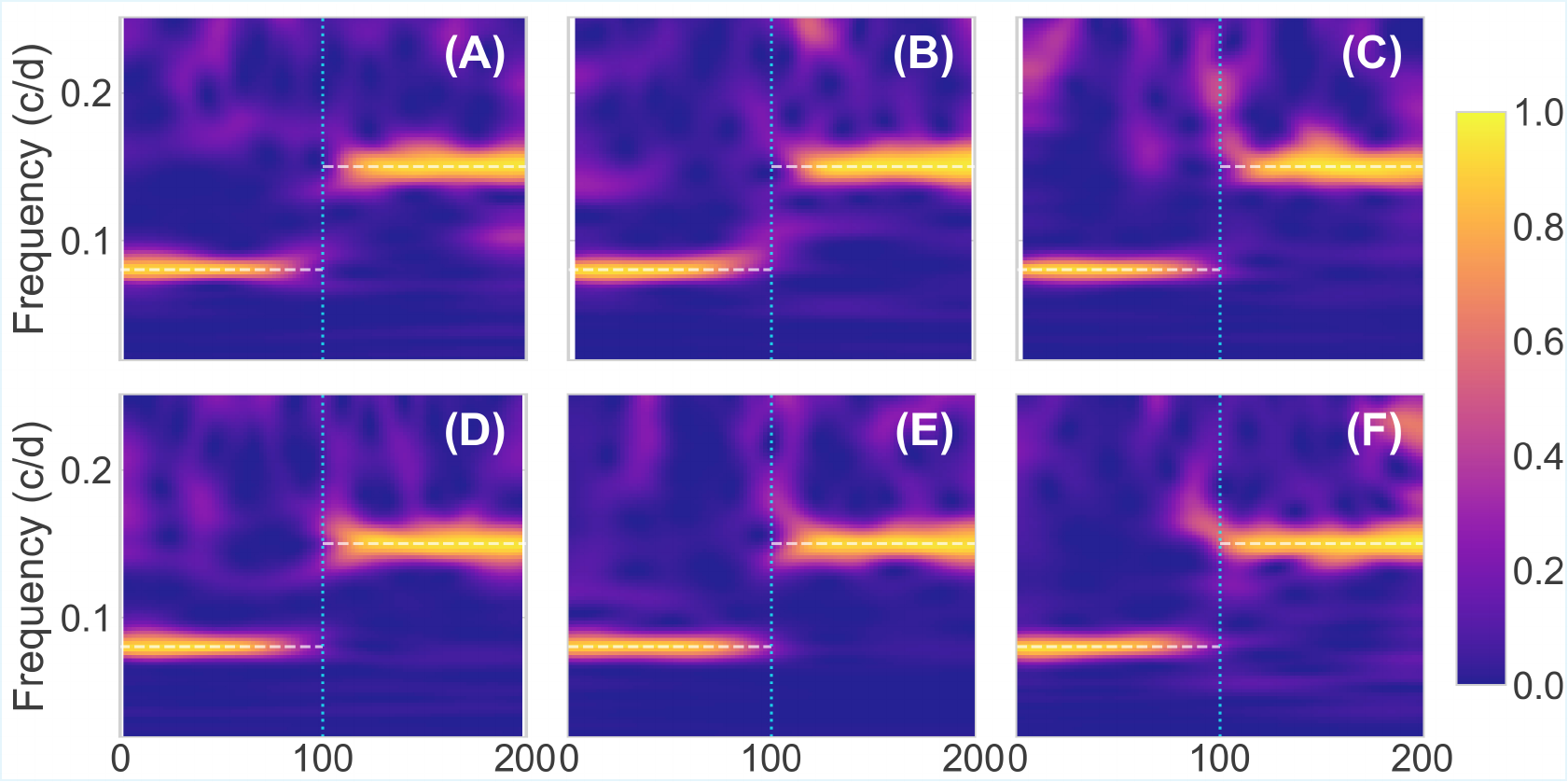}
    \setcounter{figure}{3}
    \caption{NUST spectrograms of a transient frequency jump signal obtained over multiple runs with random non-uniform sampling, while the parameters $h$, $\alpha$, and $\gamma$ are kept constant. The colour bar represents the normalised power.
}   
\vspace{-6pt}
\label{fig:rnd}
\end{figure}
The method was further validated on the public High Accuracy Radial Velocity Planet Searcher (HARPS) time series for the Sun-like star HD 10180 \cite{lovis2011harps}. Radial velocity (RV) measurements detect exoplanets by monitoring stellar line-of-sight motion from orbiting companions, producing approximately sinusoidal periodic signals, for planets in low-eccentricity orbits \cite{perryman2018exoplanet}. The dataset \cite{vizier:J/A+A/528/A112} consists of 190 measurements with instrumental uncertainties, non-uniformly sampled over approximately 6.7 years. While the system contains six confirmed planets, our analysis focuses on two representative signals (planets $\mathcal{C}$ and $\mathcal{D}$) to demonstrate the method's performance. 


The NUST spectrogram is computed over a grid of time-frequency points $(\tau, f)$ which can be chosen based on the desired resolution. 
The number of points in the time grid, $\{\tau_j\}$, determines the temporal display resolution of the spectrogram. 
For the signals in this study, a uniform grid of $100-200$ points is used for $\{\tau_j\}$.
The frequency grid $\{f_k\}$ is uniform over $[f_{\text{min}}, f_{\text{max}}]$ with $100-250$ points.
The performance of NUST is evaluated against two benchmarks: the time-integrated GLS periodogram, applied directly to the non-uniformly sampled data, and the standard S-transform, applied to an evenly sampled version of the signal with same number of samples (this serves as a reference ground-truth spectrogram). 
The implementation is available on Github \cite{githubrepo}.
\vspace{-8pt}
\subsection{Comparative analysis}
\vspace{-2pt}
A comparison of the three methods across the four signals is shown in \autoref{fig:plot4x3}. Column (A) presents the S-transform ground truth, localizing the signal features in time and frequency. Column (B) displays the GLS periodogram which averages signal power over the entire duration obscuring transient and non-stationary features. This results in ambiguous peaks for signals $1, 2,$ and $3$ that lack essential temporal and frequency information. In contrast, NUST in column (C), localizes signal features in both time and frequency, closely matching the S-transform ground truth despite being derived from non-uniformly sampled data. \par
\begin{figure}[h] 
    \centering
    \includegraphics[width=\columnwidth]{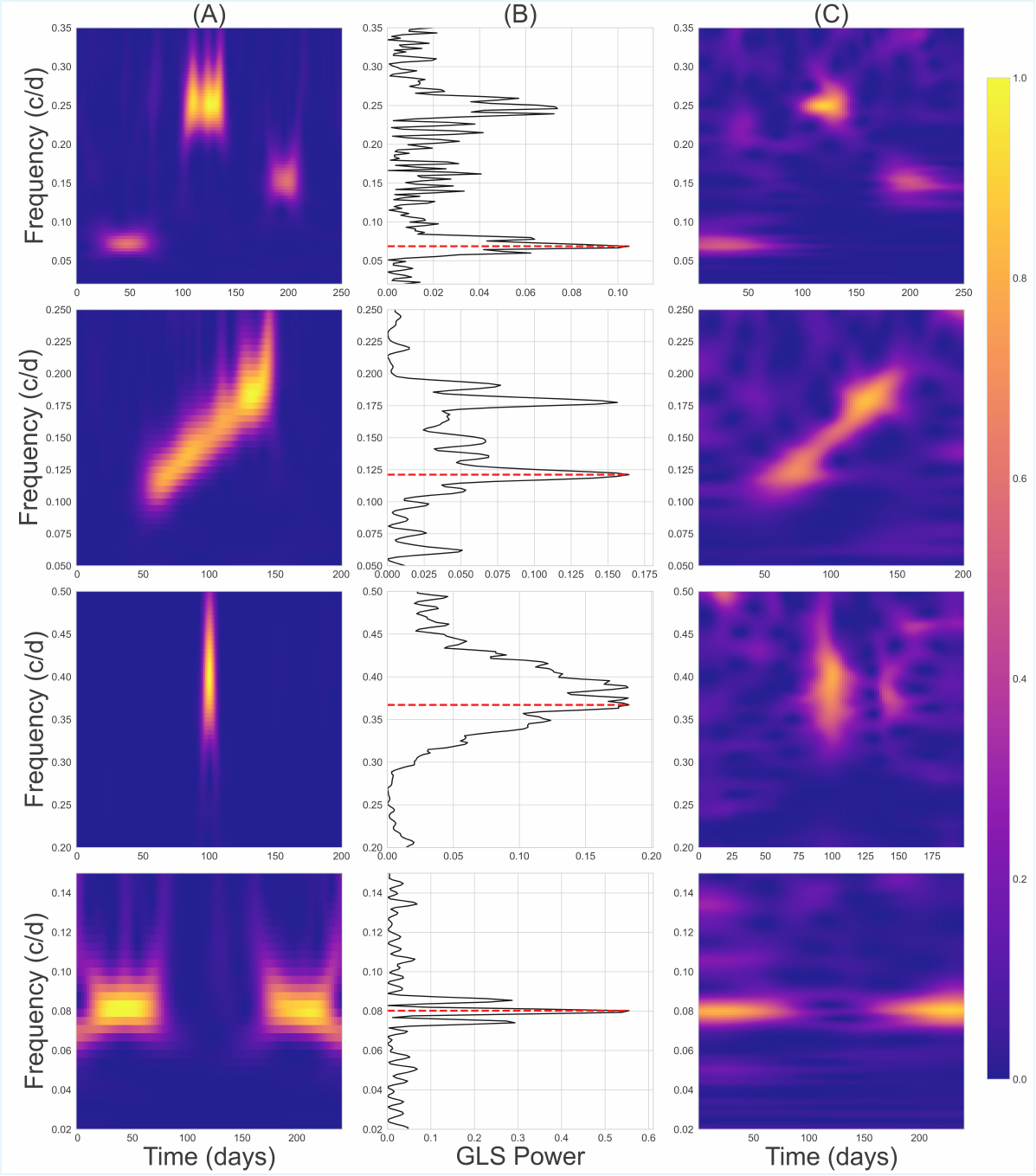}
    \setcounter{figure}{2}
    \caption{(A) The S-transform spectrograms for the 4 evenly sampled signals. (B) GLS periodograms of the corresponding non-uniformly sampled signals, with dotted red lines indicating the maximum power recovered. (C) NUST spectrograms for the same. The colour bar represents the normalised power.}
     \vspace{-8pt}
    \label{fig:plot4x3}
\end{figure}

\vspace{-6pt}
\subsection{Robustness tests}
\vspace{-2pt}
To test NUST's robustness against the specific realization of random sampling, a frequency-hopping signal was analyzed six times, each with a different random set of sample points but fixed hyperparameter values $ h, \gamma, \alpha$. As shown in \autoref{fig:rnd}, NUST consistently recovers the signal's time-frequency structure across all runs.\par
The robustness of NUST to noise was evaluated using the multi-transient signal (Signal~1) at SNR of $8.0, 2.5$, and $0.8$. The results in \autoref{fig:snr} show that NUST identifies the localised frequency information for components at high and moderate SNR. White boxes mark the approximate time interval during which the signal is expected to be present. At a low SNR of $0.8$, the signal features, especially the low frequency components, begin to merge with the noise floor, defining the practical limits of detection.
\vspace{-6pt}
\begin{figure}[h]
    \centering
    \includegraphics[width=\columnwidth]{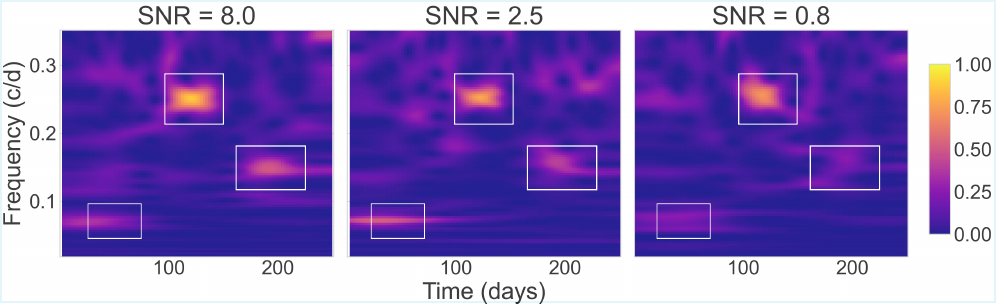}
    \setcounter{figure}{4}
    \vspace{-8pt}
    \caption{NUST spectrograms of Signal~1 under different noise conditions with SNR values of $8$, $2.5$ and $0.8$. White boxes indicate the approximate time window in which the signal is expected. The colour bar represents the normalised power.
}    
    \vspace{-10pt}
    \label{fig:snr}
\end{figure}
\subsection{HARPS RV time series analysis of HD 10180}
To demonstrate utility on real-world data, NUST was applied to the HARPS RV time series of the multi-planet system HD 10180. The analysis focused on targeted frequency ranges to test specific known planetary signals.
For the high-frequency planet $\mathcal{C}$ ($0.1736,\text{c/d}$), GLS, as seen in panel A of  \autoref{fig:hd}, shows a strong peak but also significant nearby peaks make characterization ambiguous. NUST resolves this by revealing a stable horizontal track throughout the observation period (panel B of \autoref{fig:hd}), clearly distinguishing the persistent planetary signal from transient features.

For the lower-frequency search for planet $\mathcal{D}$ ($0.061,\text{c/d}$), both GLS and NUST detect the signal as seen in panels C, D of \autoref{fig:hd}. The NUST reveals a track that, while present, appears less consistent and intermittently weaker than that of planet $\mathcal{C}$. This is an expected consequence of the uncertainty principle \cite{gabor1947theory}. To achieve fine frequency resolution for low-frequency signals, NUST uses wide analysis windows with tuned $\sigma_{\text{adaptive}}(\tau, f)$. However, excessively wide windows approach GLS behavior, loosing temporal information.
Even with this limitation, NUST’s ability to visualize the temporal behavior of signals provides a valuable diagnostic for separating coherent planetary signals from transient or quasi-periodic stellar activity.
\begin{figure}[h]
    \centering
    \includegraphics[width=\columnwidth]{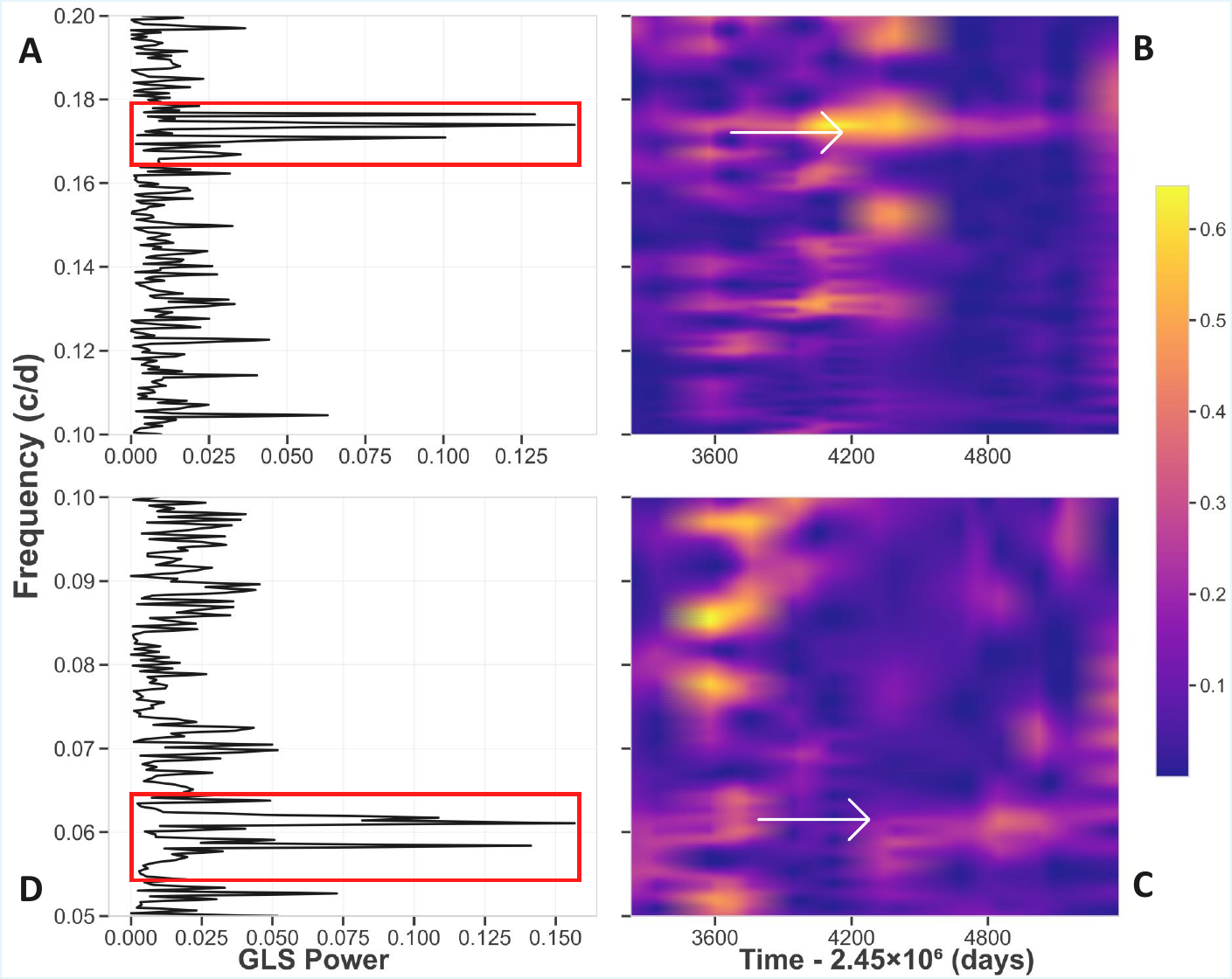}
    \setcounter{figure}{5}
    \caption{
    HD 10180: Panels A, C display the GLS periodogram for planet $\mathcal{C}$ $(f=0.1736\,\text{c/d})$ and $\mathcal{D}$ $f=0.061\,\text{c/d}$. Red boxes are placed around the true planetary peak and includes neighboring noisy peaks. Panels B, D display the NUST spectrograms. The white arrow represents the consistent frequency corresponding to the planetary signal. 
}
\vspace{-8pt}
\label{fig:hd}
    
\end{figure}

\vspace{-2pt}
\section{Conclusion}
\vspace{-2pt}
\label{sec:majhead}
We introduced the non-uniform S-transform (NUST), a time-frequency framework for non-uniformly sampled data with a doubly adaptive analysis window adjusting to both frequency and local sample density. Tests on synthetic transients, chirps, and bursts showed reliable time-frequency recovery where GLS periodogram often fails. Application to HARPS radial velocity data of HD~10180 demonstrated
isolation of persistent planetary signals from time-variable noise. 
By adding temporal context absent in the standard periodograms, NUST provides a powerful diagnostic tool.
Future work could generalize this framework to include wavelets for non-uniform data processing.

\vfill\pagebreak




\newpage
\balance
\bibliographystyle{IEEEbib}
\bibliography{strings,refs}

\end{document}